\documentclass[PRB,twocolumn,showpacs,amsmath,superscriptaddress]{revtex4-1}
\usepackage[colorlinks,urlcolor=blue,linkcolor=blue,citecolor=blue]{hyperref}
\usepackage{overpic}
\usepackage{subfigure}
\usepackage{latexsym}
\usepackage{braket}
\usepackage{amsfonts}
\usepackage{color,xcolor}
\usepackage{bm}
\usepackage{array}
\usepackage{booktabs}
\usepackage{caption}
\usepackage{multirow}

\makeatletter

\newcommand{\Rmnum}[1]{\expandafter\@slowromancap\romannumeral #1@}

\begin{document}

	\title{The $\mathbf{Z}_{2}$ topological invariants in 2D and 3D topological superconductors without time reversal symmetry}

	\author{Jinpeng Xiao}
	\affiliation{
		School of Mathematics and Physics, Jinggangshan University, Ji'an 343009, P. R. China}
	\author{Qianglin Hu}
    \affiliation{
		School of Mathematics and Physics, Jinggangshan University, Ji'an 343009, P. R. China}
	\author{Huiqiong Zeng}
    \affiliation{
		School of Eletronic and Information Engineering, Jinggangshan University, Ji'an 343009, P. R. China}
	\author{Xiaobing Luo}
   \altaffiliation{Corresponding author: xiaobingluo2013@aliyun.com}

   \affiliation{Department of Physics, Zhejiang Sci-Tech University, Hangzhou 310018, China}
	
	\date{\today}
	
	\begin{abstract}
	In theory of topological classification, the 2D topological superconductors without time reversal symmetry are characterized by Chern numbers. However, in reality, we find the Chern numbers can not reveal the whole properties of the boundary states of the topological superconductors. We figure out some particle-hole symmetry related $\mathbf{Z}_{2}$ invariants, which provide more additional information of the topological superconductors than the Chern numbers provide. With the $\mathbf{Z}_{2}$ invariant, we define weak and strong topological superconductors in 2D systems. Moreover, we explain the causes of mismatch between the Chern numbers and the numbers of boundary states in topological superconductors, and claim that the robust Majorana zero modes are characterized by the $\mathbf{Z}_{2}$ invariant rather than the Chern numbers. We also extend the $\mathbf{Z}_{2}$ invariants to 3D non-time-reversal symmetry superconductor systems including gapful and gapless situations.

	\end{abstract}
	\maketitle
	\section{Introduction}\label{Sec1}
	Over recent years, topological superconductors (TSCs) have drawn much attention in condensed matters for the novel properties of Majorana zero boundary states\cite{Kitaev2001,Read2000,Ivanov2001,Stone2006} and the promising applications in quantum computation\cite{Kitaev2003,Stern2004,Das2006,Nayak2008}. People find a lot of ways to realize and study the TSCs, including quantum dots\cite{Sau2012,Guang2016,Zhang2016,Kulesh2020,Estrada2020,Scherubl2020}, dopping topological insulators\cite{Hor2010,Wang2012,Matano2016,Greco2018,Sun2019,Das2020,Adam2021}, promxity effect between topological insulators and superconductors\cite{Fu2008,Williams2012,Xu2014,Qing2017,Trang2020,Guo2021}, and arranging magnetic atomic on the surface of an s-wave with hlical structures\cite{Martin2012,Vazifeh2013,Xiao2015,Pershoguba2016,Li2016,Howon2018,Rex2019,Choi2019,Kamlapure2021} or SOC effect\cite{Stevan2014,Rontynen2015,Heimes2015,Pawlak2016,Ruby2017,Zhang2020,Wada2021,Wang2021}.

It is well known that the topological systems can be classified into ten classes with three global symmetries including time-reversal, particle-hole and chiral symmetry\cite{Schnyder2008,Ryu2010}. The classification tells us that the different symmetries and dimensions of the systems result in different topological invariants. For example, in time-reversal symmetric TSCs, the invariants are time-reversal symmetry related $\mathbf{Z}_{2}$ invariants in 2D and 3D systems. In non-time-reversal symmetric situations, the superconductors are characterized by $\mathbb{Z}$ invariants in 2D systems and have no invariant in 3D systems. The invariants are closely related to the boundary states when imposing open boundary conditions, which provides the evidence of the robustness of the topological boundary states.

As the research of topological materials is developing, people find the invariants in the ten fold classification sometimes can not completely reflect all the aspects of the topological properties. It is observed that there are topologically nontrivial robust boundary states in the non-invariant classes, or there is no match between the invariants and the boundary states\cite{Qi2010,Zhang2013,Ueno2013,Ando2015,Shiozaki2016,Wang2016Q,Liu2017,Zhang2018,Zhang2019,Pan2019,Hu2020,Ono2021}. Lots of these anomalous phenomena appear in the spatial symmmetric systems, which can be explained by revising the classification\cite{Chiu2013,Shiozaki2014,Geier2020}. However, there still are some situations that can not be included in the revised classification\cite{Qi2010,Liu2017,Liu20172,Wu2020}. In the topological insulators, these anomalous phenomena can be explained from the perspective of non-zero 2D Zak phase\cite{Liu20172} and nonzero Berry curvature\cite{Wu2020}. In the TSC systems, the anomalous phenomenon may appear in the form of mismatchness of the Chern numbers and Majorana zero modes (MZMs). For instance, in the TSCs induced by proximity effect between topological insulators and superconductors\cite{Qi2010,Liu2017}, there are some pairs of non-zero energy boundary states inheriting from the topological insulators, which affect the Chern numbers and cause the mismatching. The mismatchness of the Chern numbers and Majorana zero modes still await comprehensive investigations.

In this work, we study the correspondence of the $\mathbf{Z}_{2}$ invariant and the MZMs in TSC without any other symmetrys except the intrinsic particle-hole symmetry (PHS). We figure out the $\mathbf{Z}_{2}$ invariants pertaining to the PHS and corresponding to the Zak phases of the lines of the Brillouin zone (BZ) boundaries. We investigate the class D systems from point of view of the $\mathbf{Z}_{2}$ invariants in 2D and 3D situations. Through definition of strong and weak TSCs in 2D systems with the $\mathbf{Z}_{2}$ invariants, we reveal that the $\mathbf{Z}_{2}$ invariants are more appropriate to characterize the properties of the MZMs than the Chern numbers. Furthermore, the explanation of the mismatchness between Chern number and MZMs is given, and the locations of the MZMs are fixed. We extend the $\mathbf{Z}_{2}$ invariant to 3D gapful and gapless systems although there is no topological invariant in the topological classification, and find that the 3D $\mathbf{Z}_{2}$ invariant provides a simple method to judge whether the gapless systems have non-degenerate Weyl nodes or not.

The paper is organized as follows. In section \ref{Sec2}, we show the relationships of the Kitaev $\mathbf{Z}_{2}$ invariant and Zak phase in 2D system, and define a 2D $\mathbf{Z}_{2}$ invariant. In section \ref{Sec3}, we discuss the causes of the mismatchness of the Chern numbers and the numbers of MZMs. In section \ref{Sec4}, we study a concrete model. In section \ref{Sec5}, we extend the $\mathbf{Z}_{2}$ to 3D situations in gapful and gapless systems, respectively. In section \ref{Sec8}, we make a brief summary.

	\section{The deduction of the $\mathbf{Z}_{2}$ invariants}\label{Sec2}
In 1D non-time-reversal symmetric TSC systems, the 1D Zak phase and Kitaev $\mathbf{Z}_{2}$ invariant are equivalent\cite{Budich2013}. In 2D systems with the same symmetries, the systems are characterized by the Chern numbers which are $\mathbb{Z}$ invariants. However, it is reasonable to predict that there are $\mathbf{Z}_{2}$ invariants in 2D systems, but the forms of the $\mathbf{Z}_{2}$ invariants and the roles of the Zak phases are yet unknown.
In a superconductor system, the Berry connection of the $n$th band is defined as $A_{n}(\mathbf{k})=i<u_{n}(\mathbf{k})|\partial_{\mathbf{k}}|u_{n}(\mathbf{k})>$, where $|u_{n}(\mathbf{k})>$ is the bloch wave function of the $n$th band. By taking advantage of the PHS, the Berry connection has the form $A_{n}(-\mathbf{k})=i<u_{n}(-\mathbf{k})|\partial_{-\mathbf{k}}|u_{n}(-\mathbf{k})>=A_{-n}(\mathbf{k})-\partial_{\mathbf{k}}\chi_{n}(\mathbf{k})$. In the superconductor system, the $-n$th band is particle-hole symmetric with the $n$th band. We have $A^{o}(-\mathbf{k})=A^{e}(\mathbf{k})-\sum_{n}\partial_{\mathbf{k}}\chi_{n}(\mathbf{k})$ for all the bands with $o$ the energies below zero energy and $e$ the energies above the zero energy.

The Chern number of the 2D TSC is related to the Berry phase of the bands below zero energy. In the square lattice model, the Berry phase of the bands below zero energy can be written as
\begin{eqnarray}\label{eq1}
\psi=\oint_{L} A^{o}(\mathbf{k})d\mathbf{k}=\sum_{i=1}^{4}\oint_{L_{i}} A^{o}(\mathbf{k})d\mathbf{k},
\end{eqnarray}
where the integral loop $L$ is the boundary of the first BZ. The loops $L_{i}$ are the boundaries of the 1/4 areas of the first BZ from the first quadrant to the fourth quadrant as shown in Fig. \ref{fig1}(a). The second equality in Eq. (\ref{eq1}) is the sum of 1D Zak phase around four lines.
In 1D TSC, the Zak phase becomes $\psi_{Zak}=\int^{\pi}_{-\pi}A^{o}(k)dk=\int^{\pi}_{0}A^{o}(k)dk+\int^{0}_{-\pi}A^{o}(k)dk=\int^{\pi}_{0}A(k)-\sum_{n}\chi_{n}(k)dk$ with $A(k)=A^{o}(k)+A^{e}(k)$. In the system with PHS, we can choose the gauge which makes $\chi_{n}(k)=0$, and then the Berry phase (or 1D boundary Zak phase) becomes
\begin{eqnarray}\label{eq2}
\psi&=&\oint_{L_{1}} A(\mathbf{k})dk+\oint_{L_{3}} A(\mathbf{k})dk,
\end{eqnarray}
with routes $L_{1},L_{3}$ being the boundaries of the 1/4 BZ in first quadrant and third quadrant, respectively.

In a SC system, the Hamitonian can be written in the Majorana picture as
\begin{eqnarray}\label{eq3}
H=\frac{i}{2}\gamma^{\dag}B_{m}\gamma,
\end{eqnarray}
where $B_{m}$ is a real anti-symmertric matrix. There is a real-valued orthogonal transformation matrix $W$ which can block diagonalize the $B_{m}$ as $B_{d}=W^{\dag}B_{m}W=diag_{\lambda}i\varepsilon_{\lambda}\sigma_{y}$. With Fourier transformation, $B_{d}(k)=W(k)^{\dag}B_{m}(k)W(k)=diag_{\lambda}i\varepsilon_{\lambda}(k)\sigma_{y}$. And with a $k$-independent unitary transformation matrix $U$, $B_{d}(k)$ can be diagonalized as $U^{\dag}B_{d}(k)U=diag (i\varepsilon_{1}(k),...,i\varepsilon_{n}(k))$. So we have $U^{\dag}W(k)^{\dag}iB_{m}(k)W(k)U=diag_{n}(\varepsilon_{n}(k))$, and the transformation matrix $W(k)U$ is composed of bloch functions $|u_{n}(k)>$. Since the $k$-independent unitary transformation does not change the trace of a matrix, the Berry phase can be calculated as
\begin{eqnarray}\label{eq4}
\psi&=&i\oint_{L_{1}+L_{3}}Tr[W(\mathbf{k})^{\dag}\partial_{\mathbf{k}}W(\mathbf{k})]d\mathbf{k}\nonumber\\
&=&i\oint_{L_{1}+L_{3}}\partial_{\mathbf{k}}ln det [W(\mathbf{k})]d\mathbf{k}.
\end{eqnarray}
The phase $\chi_{n}(\mathbf{k})$ is zero in this choosing of gauge in that the PHS requires $CW(\mathbf{k})C^{-1}=W(-\mathbf{k})$.

The block diagonalized Hamitonian matrix $B_{d}$ in Majorana picture is also anti-symmetryic, and we observe that the Pfaffian of $B_{d}$ is positive forever, which can be calculated as
$Pf[B_{d}]=Pf[W^{\dag}B_{m}W]=Pf[B_{m}]det(W)>$1 and gives $Pf[B_{m}]=det(W)=\pm 1$. The fact that Fourier transformation does not change the Pfaffian of a system guarantees $Pf[B_{m}]=\prod_{\mathbf{k}}Pf[B_{m}(\mathbf{k})]=\prod_{\mathbf{k}}det[W(\mathbf{k})]$ to hold. $W(\mathbf{k})$ is unitary, so $det[W(\mathbf{k})]=e^{i\Phi_{\mathbf{k}}}$. Because the matrix $W$ is real, we have $W(\mathbf{k})^{*}=W(-\mathbf{k})$ and $det[W(\mathbf{k})]^{*}=det[W(-\mathbf{k})]$, and we get $e^{i\Phi_{\mathbf{k}}}=e^{-i\Phi_{-\mathbf{k}}}$, which results in $\Phi_{\mathbf{k}}=-\Phi_{-\mathbf{k}}$ mod $2\pi$. As any $det[W(\mathbf{k})]$ in $k$ is identical to that in $-k$, we have $\prod_{\mathbf{k}}det[W(\mathbf{k})]=\prod_{\mathbf{K}}det[W(\mathbf{K})]$ with $\mathbf{K}$ the particle-hole symmetric points, and $Pf[B_{m}]=\exp(i\sum\epsilon_{\mathbf{K}_{j}}\Phi_{\mathbf{K}_{j}})$ with $\epsilon_{K_{j}}=\pm 1$. In 2D system, we take $K_{1},K_{2},K_{3},K_{4}=(0,0),(0,\pi),(\pi,\pi)$,$(\pi,0)$, or take $K_{1},K_{2},K_{3},K_{4}=(-\pi,-\pi),(-\pi,0),(0,0)$,$(0,-\pi)$ because of the translation invariance, thus leading to
\begin{eqnarray}\label{eq5}
Pf[B_{m}]=\exp(i\phi_{j}),
\end{eqnarray}
with $j=1,2,3,4$, $\phi_{1}=\int_{0}^{\pi}\partial_{k_{x}}\Phi(k_{x},\pi)dk_{x}+\int_{\pi}^{0}\partial_{k_{x}}\Phi(k_{x},0)dk_{x}$,
$\phi_{2}=\int_{0}^{\pi}\partial_{k_{y}}\Phi(0,k_{y})dk_{y}+\int_{\pi}^{0}\partial_{k_{y}}\Phi(\pi,k_{y})dk_{y}$,
$\phi_{3}=\int_{-\pi}^{0}\partial_{k_{x}}\Phi(k_{x},0)dk_{x}+\int_{0}^{-\pi}\partial_{k_{x}}\Phi(k_{x},-\pi)dk_{x}$,
$\phi_{4}=\int_{-\pi}^{0}\partial_{k_{y}}\Phi(-\pi,k_{y})dk_{y}+\int_{0}^{-\pi}\partial_{k_{y}}\Phi(0,k_{y})dk_{y}$
. Here $\phi_{1}$(mod $2\pi$)=$\phi_{2}$(mod $2\pi$)=$\phi_{3}$(mod $2\pi$)=$\phi_{4}$(mod $2\pi$). According to the determinant of the transformation matrix $W(\mathbf{k})$, we have
\begin{eqnarray}\label{eq6}
\phi_{1}&=&i\int_{l_{1}+l_{3}}\partial_{\mathbf{k}}lndet[W(\mathbf{k})]d\mathbf{k},\nonumber\\
\phi_{2}&=&i\int_{l_{2}+l_{4}}\partial_{\mathbf{k}}lndet[W(\mathbf{k})]d\mathbf{k},\nonumber\\
\phi_{3}&=&i\int_{l'_{1}+l'_{3}}\partial_{\mathbf{k}}lndet[W(\mathbf{k})]d\mathbf{k},\nonumber\\
\phi_{4}&=&i\int_{l'_{2}+l'_{4}}\partial_{\mathbf{k}}lndet[W(\mathbf{k})]d\mathbf{k},
\end{eqnarray}
with $l_{i}(l'_{i})(i=1,2,3,4)$ being the four sides of $L_{1}(L_{3})$ as shown in Fig. \ref{fig1}(a). In combination with Eq. (\ref{eq2}) and Eq. (\ref{eq4}), we obtain $\phi_{j}=\int_{l_{s}}A(\mathbf{k})d\mathbf{k}$ with $l_{s}$ being the two counter lines, and the Berry phase $\psi=\sum_{j=1}^{4}\phi_{j}$. It follows that
\begin{eqnarray}\label{eq7}
\exp(i\psi)=Pf^{4}[B_{m}]
\end{eqnarray}
is definitely +1.
Therefore, the Berry phase can't be directly used to characterize the $\mathbf{Z_{2}}$ invariant in the 2D systems. However, the discussion above provides an alternative way to define a $\mathbf{Z_{2}}$ invariant.
For $\phi_{j}=0,\pi$ mod 2$\pi$, which are the same in loops $L_{1}$ and $L_{3}$, we can define the $\mathbf{Z_{2}}$ invariant as
\begin{eqnarray}\label{eq8}
M_{2D}=(-1)^{\phi_{j}/\pi},
\end{eqnarray}
which is equivalent to the Pfaffian of $B_{m}$. There are
$M_{2D}=Pf(B_{m})=Sgn\{\prod_{K_{i}}Pf[B_{m}(K_{i})]\}$
with $K_{i}$ the particle-hole symmetric invariant points in the first BZ.

Furthermore, each $\phi_{j}$ consists of two parts, which both are equal to 0 or $\pi$ mod 2$\pi$. Taking $\phi_{1}$ for example, the two parts are given by $\int_{l_{1}}\partial_{\mathbf{k}}\Phi(\mathbf{k})d\mathbf{k}=i\int_{l_{1}}\partial_{\mathbf{k}}lndet[W(\mathbf{k})]d\mathbf{k}$=0 or $\pi$ mod 2$\pi$ and $\int_{l_{3}}\partial_{\mathbf{k}}\Phi(\mathbf{k})d\mathbf{k}=i\int_{l_{3}}\partial_{\mathbf{k}}lndet[W(\mathbf{k})]d\mathbf{k}$=0 or $\pi$ mod 2$\pi$. Therefore, we can define the sub-$\mathbf{Z_{2}}$ invariants under $M_{2D}$ as
\begin{eqnarray}\label{eq9}
M_{j}=(-1)^{i\int_{l_{j}}\partial_{\mathbf{k}}\Phi(\mathbf{k})d\mathbf{k}}=(-1)^{i\int_{l_{j}}A(\mathbf{k})d\mathbf{k}},
\end{eqnarray}
with $j=1,2,3,4$.
It is not hard to see
$M_{j}=Sgn\{\prod_{i=j-1}^{j}Pf[B_{m}(K_{i})]\}$, which are just the 1D Kitaev $\mathbf{Z}_{2}$ invariants in the lines $l_{j}$.
Actually, we can know all the four $M_{j}$ as long as we know any neighboring two of them. Therefore, the whole $\mathbf{Z_{2}}$ invariant of a 2D TSC can be denoted as $(M_{2D};M_{1},M_{2})$ with $M_{1},M_{2}$ associated with any two neighbor lines. In this paper, we shall take the two neighbor lines around the coordinate origin (i.e., line $l_{1}$, $l_{2}$ in Fig. \ref{fig1}(a)) as the integral lines of $M_{1}$ and $M_{2}$, respectively.

	\section{The mismatch of the Chern number, $\mathbf{Z_{2}}$ topological invariant and the edge states}\label{Sec3}
\begin{figure}
\scalebox{1.0}{\includegraphics[width=0.5\textwidth]{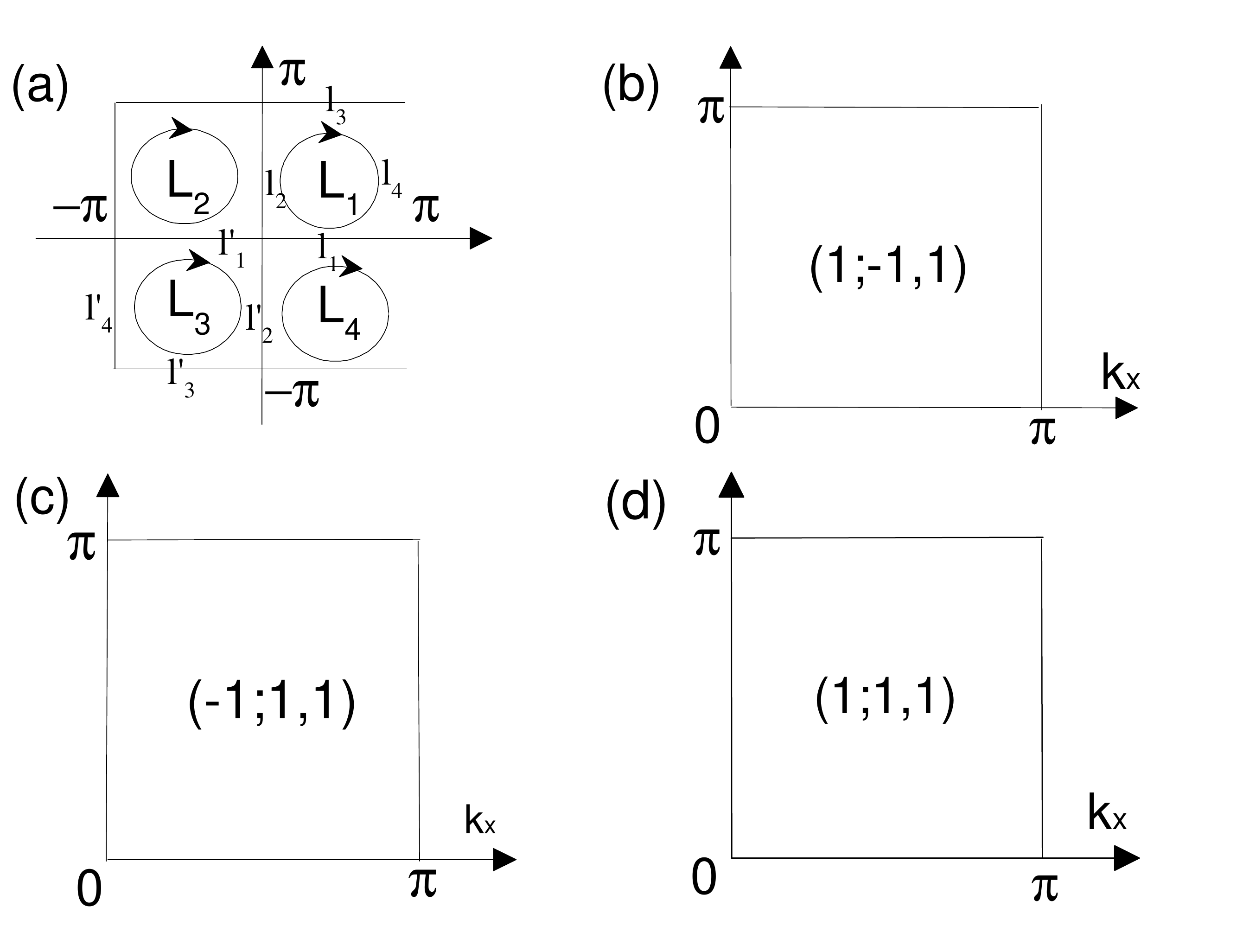}}
\caption{\label{fig1}  The BZ of 2D systems and the $\mathbf{Z}_{2}$ invariants. (a) shows the 2D BZ with the integral loops in Eq. (\ref{eq1}) from $L_{1}$ to $L_{4}$, and all the directions of line integrals are clockwise. (b)-(d) show the situations with Chern number $C=0,1$ and $2$, respectively. The $\mathbf{Z}_{2}$ invariants are denoted in the form of $(M_{2D};M_{1},M_{2})$ in (b)-(d). (b) and (c) are also capable of describing the situations with Chern numbers being $2n$ and $2n+1$ respectively. Here $n=0,\pm1,\pm2,....$}
\end{figure}

In general, the class A (the system without any symmetry) and the class D systems are characterized by Chern numbers in 2D situation, and the numbers of boundary states are related to the Chern numbers. However, the additional PHS in class D (not in class A) provides more information about the topological properties which can not be obtained from the Chern numbers but only from the $\mathbf{Z}_{2}$ topological invariant we propose here. For instance, one can not find the locations of the MZMs from the Chern numbers, but can find them from the $\mathbf{Z}_{2}$ invariant. The signs of $M_{i}$ suggest the locations of the PHS protected MZMs as shown in Fig. \ref{fig1}(b)(c).

In addition, the Chern numbers sometimes mismatch with the numbers of MZMs.
There are three distinct situations for the mismatches which are listed as follows. First, there are MZMs in the systems, but there is no connection between the MZMs and the Chern numbers. The phenomenon can be seen in Fig. \ref{fig1}(b), where we exhibit schematic representation of 1/4 area of a 2D BZ of a class D system including four particle-hole symmetric points. In this case, the Chern number is zero and the $\mathbf{Z}_{2}$ invariants are ($1;-1,1$) as indicated in the figure. There exist robust topological MZMs locating at $k_{y}=0$ and $k_{y}=\pi$ when the open boundary conditions are applied for $x$ direction. These unique properties can be explained from perspective of one-dimensional systems. At the lines $k_{x}=0(k_{y}=0)$ and $k_{x}=\pi (k_{y}=\pi)$, the system still have PHS which makes it being 1D class D SC at the particle-hole symmetric lines. Here the Kitaev $\mathbf{Z}_{2}$ invariants are still valid, which are just the invariants defined as $M_{j}$ in Eq. (\ref{eq9}). Due to $M_{1}(k_{y}=0)=Sgn\{Pf[B_{m}(0,0)]Pf[B_{m}(\pi,0)]\}=-1$ and $M_{3}(k_{y}=\pi)=Sgn\{Pf[B_{m}(0,\pi)]Pf[B_{m}(\pi,\pi)]\}=-1$, the lines $k_{y}=0$ and $k_{y}=\pi$ are topologically nontrivial. However, at lines $k_{x}=0$ and $k_{x}=\pi$, we have $M_{2}=M_{4}=+1$ and thus the two lines are trivial, having no robust topological MZMs when the open boundary conditions in $y$ direction are applied. In such situations, the MZMs are protected by PHS rather than the Chern numbers. The same description can be applied in the systems, where the Chern numbers are 2$n$, and nevertheless, the topological edge states are 2$n$+2 with two of them being MZMs by imposing open boundary conditions. In these systems, there are odd number (i.e., $2m+1$) of edge states locating at the ends of line $k_{y}=0$ and odd number (i.e., $2l+1$) locating at the ends of line $k_{y}=\pi$ as well. The other left edge states (i.e., $2(n-m-l)$) locate at $k_{y}$ and $-k_{y}$ in pairs. However, it should be noted that not all the edge states are symmetry protected MZMs. The edge states locating at the $k_{y}$ and $-k_{y}$ ($k_{y}\neq 0,\pi$) may feature zero energy, but they are not robust. Even at the particle-hole symmetric positions $k_{y}=0,\pi$, pairs of MZMs may annihilate when the numbers of edge states at the same place are more than 1. So, in the square lattice systems, there are at most two PHS protected MZMs locating at the two particle-hole symmetric positions when we open the boundaries of one direction. The other zero modes are not protected by any symmetry, and are fragile to perturbations.

The second situation is that the Chern numbers are nonzero but are partially related to the MZMs. The partial connection is not so obvious in the simple case where Chern number and the number of MZMs are both equal to one as shown in Fig. \ref{fig1}(c). But it is obvious in the high Chern number situations. When the Chern number is $2n+1$, there is only one robust MZM locating at the particle-hole symmetric position which can be fixed by the $\mathbf{Z}_{2}$ invariants $M_{j}$. The other edge states behave in the same way as discussed above: pairs of them locate at the $k_{i}$ and $-k_{i}$, and some other pairs locate at the particle-hole symmetric positions. Among these edge states, the MZMs corresponding to zero energies are nonetheless vulnerable. One of the proper explanations is given by the inheriting theory\cite{Liu2017}. Let us start with the normal states Hamitonian of the TSC. If the Chern number of the normal states Hamitonian is non-zero, the Chern number will be doubled when adding the SC pairing term in the Hamiltonian within BdG mean field approximation. The reason is that the order of the normal Hamitonian matrix is doubled in the process. If the pairing terms don't totally destroy the topological structures of the normal bands, there will exist pairs of edge states which are not necessarily robust MZMs. These edge states are inherited from the normal Hamitonian. However, due to the SC pairing, another non-zero Chern number may come in, which is related to the robust MZMs. In such situation, given the existence of different origins of the Chern number, we can not acquire the detailed properties of the MZMs through the Chern numbers.

The last situation is that the Chern numbers are nonzero but the number of MZMs is zero. Here, the Chern numbers merely give the numbers of the edge states which are not robust MZMs at all. In Fig. \ref{fig1}(d), we show the situation where the Chern number is two, while the number of robust MZMs is zero. The general cases with even Chern numbers but without robust MZMs can also be represented by the same schematic as Fig. \ref{fig1}(d). In such systems, $M_{2D}$ and all the $M_{j}$ are +1. The trivial $\mathbf{Z}_{2}$ invariant and vanishing robust MZMs indicate that the systems are not TSCs. The cause of the phenomenon can be also explained by the inheriting theory. The only difference from the second situation is that the SC pairing terms don't bring odd Chern numbers.

We conclude that the $M_{2D}$ is equivalent to the parity of the Chern number. Specifically, there are two scenarios. $Case$ 1: The Chern number is odd. In this case, according to the PHS, there must be odd numbers of edge states locating at the particle-hole symmetric points when opening boundaries along one direction. Likewise, when we open boundaries along $x$ direction, there must be odd numbers of zero modes locating at $k_{y}=0$ or $\pi$. Therefore, we derive $\phi_{j}=\pi$ mod $2\pi$. $Case$ 2: The Chern number is even. The PHS ensures even numbers (possibly zero) of the edge states locating at the particle-hole symmetric points with open boundaries along one direction. In both cases, $\phi_{j}=0$ mod $2\pi$, such that
\begin{eqnarray}\label{eq10}
M_{2D}=(-1)^{\psi/2\pi}=(-1)^{C},
\end{eqnarray}
with $C$ being the Chern number of the system.

The discussion above tell us that the MZMs in TSCs are protected by the PHS and characterized by the $\mathbf{Z}_{2}$ invariants rather than the Chern numbers. The Chern numbers only tell the numbers of boundary states, while the $\mathbf{Z}_{2}$ invariants tell the numbers and locations of the MZMs. We study TSC here in the similar manner as people conventionally treat the boundary states in the time-reversal symmetric topological insulators\cite{Fu2007} which are also characterized by $\mathbf{Z}_{2}$ invariants. We classify the systems as strong and weak TSCs in the same fashion. In $(M_{2D}; M_{1},M_{2})$, the strong index is given by $M_{2D}$ and the weak indices are $M_{j}$. When $M_{2D}=-1$, there must be one robust edge MZM when open boundary conditions in any one direction is applied. The other even numbers of edge states locate at arbitrary $k$ and $-k$ positions in pairs. We call such systems strong topological superconductors (STSCs). When $M_{2D}=+1$, and either of $M_{i}$ or both of $M_{i}$ are -1, we call the systems weak topological superconductors (WTSCs). In such cases, there must be two robust edge MZMs at any open boundary direction when both $M_{1}$ and $M_{2}$ are -1, or two robust edge MZMs at only one open boundary direction when only one of $M_{1}$ and $M_{2}$ is -1. The two MZMs will annihilate each other when they adiabatically move to the same place. In summary, the MZMs in both STSCs and WTSCs are robust to particle-hole symmetric perturbations, whereas the MZMs in the WTSCs are fragile when they move to the same place.

	\section{The model Hamitonians}\label{Sec4}
We start from a trilayer spinless $p$-wave 2D superconductor. The Hamitonian in $k$ space is
\begin{eqnarray}\label{eq11}
H_{k_{x},k_{y}}&=&\varepsilon_{k}\sigma_{z}+V\tau_{z}\Gamma_{z}+t_{z}\Gamma_{x}\sigma_{z}\nonumber\\
&+&\Delta\sin k_{x}\Gamma_{z}\sigma_{x}+\Delta\sin k_{y}\Gamma_{z}\sigma_{y},
\end{eqnarray}	
where $\varepsilon_{k}=\mu+2t_{x}\cos k_{x}+2t_{y}\cos k_{y}$, $\Delta, \mu, V, t_{x}, t_{y}, t_{z}$ are the order parameter, chemical potential, on site energy, hoping amplitude at $x, y$ and interlayer directions, respectively. The Pauli matrices $\sigma_{i}$ act on the particle-hole space, and $\Gamma_{z}=diag\{1,-1,1\}$, $\Gamma_{x}=$
\begin{math}
\left(
\begin{smallmatrix}
&1&\\
1&&1\\
&1&
\end{smallmatrix}
\right),
\end{math}
$\Gamma_{y}=$
\begin{math}
\left(
\begin{smallmatrix}
&&-i\\
&0&\\
i&&
\end{smallmatrix}
\right)
\end{math}
act on sublayer space. The signs of the $p$-wave pairings as well as the on-site energies in the middle layer are different from those in the first and third layers. The PHS operator is $C=i\sigma_{x}K$ with $K$ the conjugate operator.
\begin{figure*}
\centering
\scalebox{1.0}{\includegraphics[width=0.85\textwidth]{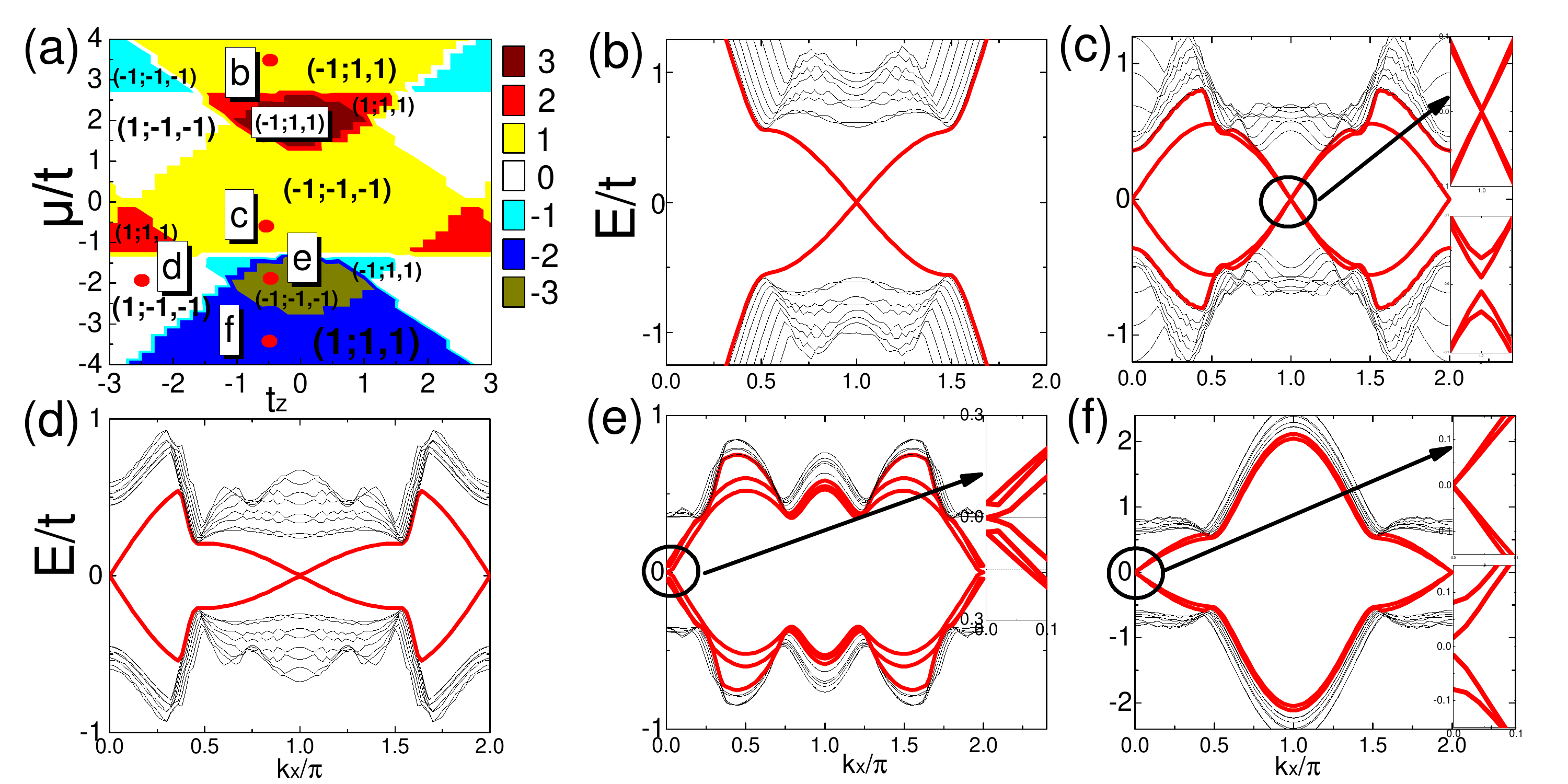}}
\caption{\label{fig2} The phase diagrams and open boundary spectrums of model (\ref{eq11}). (a) shows the phase diagrams of Hamitonian (\ref{eq11}). The colors represent different Chern numbers. Numbers in the brackets denote the strong and weak indices of the $\mathbf{Z}_{2}$ invariants. (c)-(f) show the spectrums when the parameters are taken as the letters c-h denoted in (a) with open boundary conditions. The insets in (c), (e) and (f) are the enlarged views of the cycles. The lower insets in (c) and (f) shows the open boundary spectrums with perturbations ($0.1\Gamma_{y}$) obeying PHS, and they indicate that the zero modes are distroyed. Here $t_{x}=t_{y}=t$, $V=1.3t$, $\Delta=0.6t$.}
\end{figure*}

We show the topological phase diagrams of the model in Fig. \ref{fig2}(a), where the colors refer to different Chern numbers, and the numbers in brackets refer to the $\mathbf{Z}_{2}$ invariants of the corresponding areas. It is obvious that the invariant $M_{2D}$ is in coincidence with the parity of the Chern numbers. We also show dispersions of different topological phases in Fig. \ref{fig2}(b)-(f) with the open boundary coditions in $x$ direction. In Fig. \ref{fig2}(b) and (c) with the parameters denoted as dots b and c in Fig. \ref{fig2}(a), we witness MZMs that are STSCs with identical Chern numbers. But Fig. \ref{fig2}(b) and (c) showcase different $M_{1}$ and $M_{2}$, which indicate different locations of the robust MZMs. In Fig. \ref{fig2}(b), the robust MZM locates at $k_{x}=\pi$ and in (c), the robust MZM locates at $k_{x}=0$. As shown in Fig. \ref{fig2}(c), it seems that there exist another two MZMs locating at $k_{x}=\pi$, which actually are not robust because a particle-hole symmetric perturbation may destroy them (see the lower inset of Fig. \ref{fig2}(c)). So, the Chern numbers and $M_{2D}$ do not completely provide the whole information of the boundary states. Next, we will show the mismatching of Chern numbers and MZMs.

We first show the situations that Chern numbers are zero but there exist robust MZMs.
In the white areas of Fig. \ref{fig2}(a), the Chern numbers are zero, and the strong TSC invariants are $M_{2D}=+1$. It seems like the white areas are topologically trivial. However, there exist topologically nontrivial boundary states in the systems as shown in Fig. \ref{fig2}(d) when imposing open boundary conditions. From the values of invariants denoted in the diagrams, we know that the white areas are in WTSCs phase. For example, in the white area where the dot d locates, we see $M_{1}=M_{2}$=-1, which means the line $k_{x}=0(k_{y}=0)$ and $k_{x}=\pi(k_{y}=\pi)$ are topologically nontrivial. There are two MZMs locating at $k_{x}=0(k_{y}=0)$ and $k_{x}=\pi(k_{y}=\pi)$ when we open $y(x)$ direction, respectively. The same phenomena can happen in other white areas, where the MZMs are found to be completely disconnected with the Chern numbers.

We can also find the situations that the robust MZMs are partially connecting with the Chern numbers. In the areas with Chern number being $\pm 3$ in Fig. \ref{fig2}(a), there are three topologically nontrivial boundary states. However, we see that the $M_{2D}$ equals -1 in such areas, which indicates that only one of the three is robust MZM. The boundary states of the situation with Chern number being $-3$ are shown in Fig. \ref{fig2}(e), where we observe one MZM and two topological-insulator-like boundary states locating at $k_{x}(k_{y})=0$ as $M_{1}=M_{2}=-1$ implies. In such situations, there may be many zero modes and the numbers of zero modes are determined by the absolute value of Chern numbers, but only
one of them is robust to particle-hole symmetric perturbation.

Finally, there are situations that the Chern numbers are nonzero while the numbers of robust MZMs are zero. We denote them with blue (Chern number is -2) and red colors (Chern number is 2) in Fig. \ref{fig2}(a). In these two areas, $M_{2D}$, $M_{1}$ and $M_{2}$ are all positive, which indicates that there is no PHS protected MZMs in open boundary systems. The edge states of dot f are shown in Fig. \ref{fig2}(f), where two edge states locate at $k_{x}=0$, but neither of them are robust MZMs. A particle-hole symmetric perturbation may destroy them as we show in the low inset in Fig. \ref{fig2}(f). Such situation can be viewed as the adiabatic evolution from the WTSC situation with Chern number being -2 and one of $M_{1},M_{2}$ being -1. By moving the MZM from $k_{x}=\pi$ to $k_{x}=0$, the two MZMs annihilate each other. That is why we call the TSC with $M_{2D}=+1$ as WTSC.

It is not hard to extend the discussion above to the high Chern number situations. When the Chern numbers are even (including zero), all of the boundary states are not robust MZMs, or only two of them are robust MZMs which nevertheless become fragile when moving them to the same place (WTSC). When Chern numbers are odd, only one of boundary states is robust MZM (STSC), which are immune to any particle-hole symmetric perturbations.

	\section{Topological phenomena in 3D superconductors without time reversal symmetry}\label{Sec5}
\subsection{Gapful systems}\label{Sec6}
\begin{figure}
\scalebox{1.0}{\includegraphics[width=0.45\textwidth]{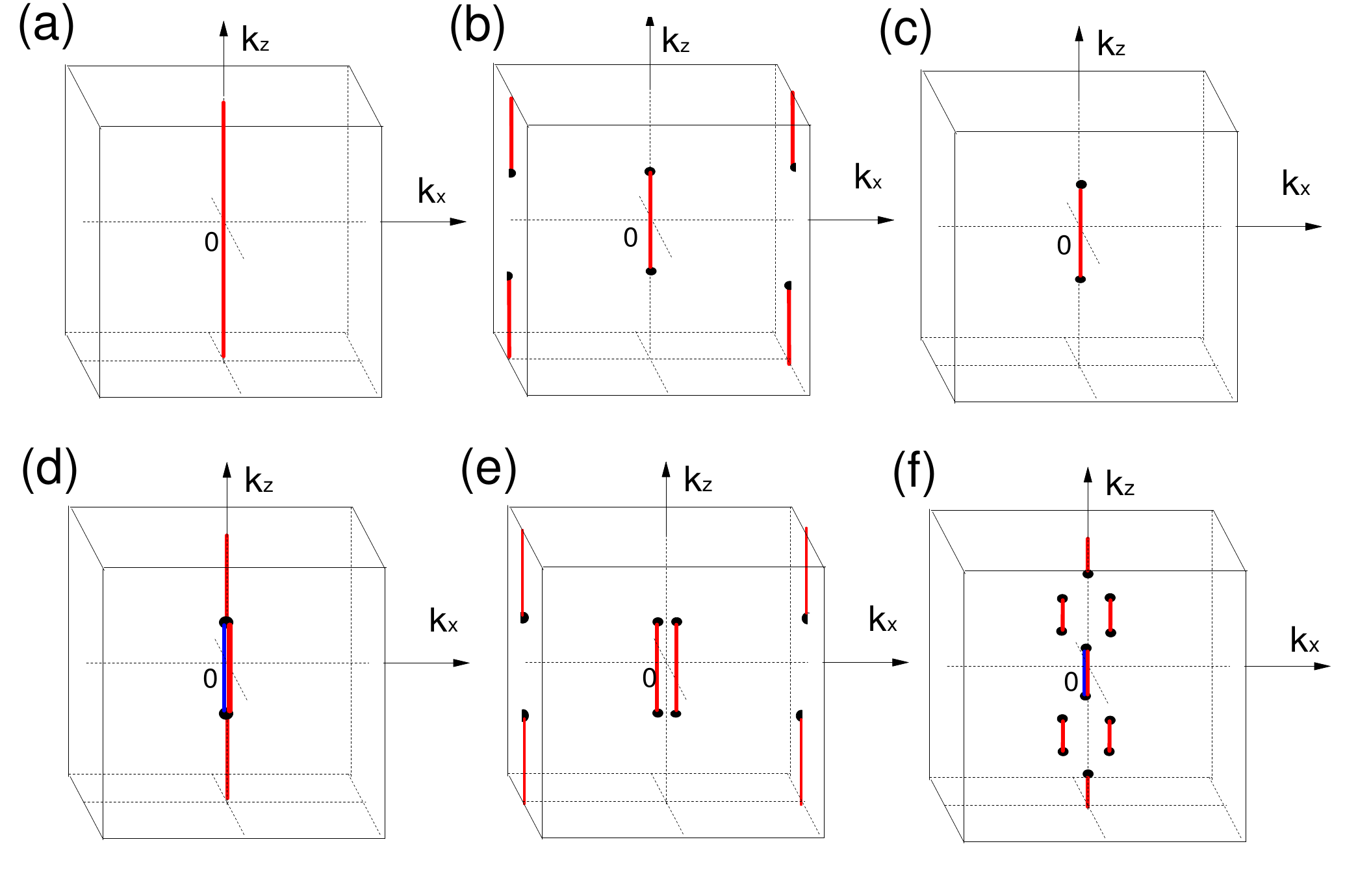}}
\caption{\label{fig3} The 3D BZ of cubic lattice model and zero modes and Weyl nodes in 3D superconductors. (a) shows the 3D BZ of gapful system and the red line is for zero mode with the open boundary condition in y direction. The nodes and the boundary states of gapless system are shown in (b)-(f). In (b) and (c), the planes $k_{z}=0$ are both STSCs, and the planes $k_{z}=\pi$ are STSC and trivial, respectively; In (d)-(f), the planes $k_{z}=0$ are WTSCs and the planes $k_{z}=\pi$ are STSCs. The black points denote the positions where the gaps close or open. The points in (d) are Dirac points, and in (f), there exist Dirac points and Weyl points. All the points in (b), (c) and (e) are Weyl points.}
\end{figure}

In 3D SC non-time reversal symmetric systems, there is no topological invariant in the topological classification table without additional symmetries\cite{Schnyder2008,Ryu2010}. But there exist many topological phenomena in the 3D systems. It is still reasonable to identify the 3D SCs with the $\mathbf{Z}_{2}$ invariant. In a certain $k$ plane of the 3D gapful system, the Hamitonian becomes two dimensional, while in $k=0$ or $\pi$ plane, the system remains particle-hole symmetric. We can use the invariant $(M_{2D};M_{1},M_{2})$ as in the 2D situations. As the system is gapful, the topological phase of plane $k=0$ must be the same as plane $k=\pi$. It comes from the fact that if the two planes are not topologically equivalent, there must be topological phase transition accompanied with band closing between the two planes. Therefore, if we know the $\mathbf{Z}_{2}$ invariant of such one plane, we naturally know the whole system's topological properties.
For example, if the plane $k_{z}=0$ is weak 2D TSC, we can find two numbers of flat bands linking the symmetry protected zero modes in the $k_{z}=0$ and $k_{z}=\pi$ when we open the boundary at $x$ or $y$ direction. If plane $k_{z}=0$ is a strong 2D TSC, the number of the flat bands is only one.
In the following, we refer to one concrete mode which differs from the one discussed in Sec. \ref{Sec4} by only stacking bilayer 2D SC (with layer dependent SC pairings and on-site energies) in $z$ direction and imposing periodic boundary conditions in $z$ direction. Thus, the Hamitonian becomes
\begin{eqnarray}\label{eq13}
H(\textbf{k})&=\Delta\sin k_{x}\tau_{z}\sigma_{x}+\Delta\sin k_{y}\tau_{z}\sigma_{y}+V\tau_{z}\sigma_{z}\nonumber\\
&+\varepsilon_{k}\sigma_{z}+t_{z}\sigma_{z}[(1+\cos k_{z})\tau_{x}+\sin k_{z}\tau_{y}],
\end{eqnarray}	
with Pauli matrices $\tau_{i}$ acting on sublayer space.
For simplicity, we choose $t_{x}=t_{y}=1$, then the Pfaffian of the four high symmetric points in $k_{z}=0$ plane read
\begin{eqnarray}\label{eq14}
&&Pf[B_{m}(0,0)]=(\mu+2)^{2}-V^{2}-4t_{z}^{2},\nonumber \\
&&Pf[B_{m}(0,\pi)]=Pf[B_{m}(\pi,0)]=\mu^{2}-V^{2}-4t_{z}^{2},\nonumber\\
&&Pf[B_{m}(\pi,\pi)]=(\mu-2)^{2}-V^{2}-4t_{z}^{2}.
\end{eqnarray}	
In the pariticle-hole symmetric invariant points, the Pfaffian of point $(0,\pi)$ is identical with $(\pi,0)$. In gapful system, plane $k_{z}=0$ and $k_{z}=\pi$ must be in the same topological phase, so here we only need to list the situation of plane $k_{z}=0$.
If the $k_{z}=0(k_{z}=\pi)$ plane is topologically nontrivial, the system is topologically nontrivial, and the topological flat bands link the zero modes in plane $k_{z}=0$ and $k_{z}=\pi$ when we open $x$ or $y$ direction. We show the STSC situation in Fig. \ref{fig3}(a) where we observe one flat band linking MZMs in planes $k_{z}=0$ and $k_{z}=\pi$. When the counter two planes $k_{z}=0$ and $k_{z}=\pi$ are in the same WTSC phase, the system can still be gapful, and there will be two flat bands linking the MZMs in the two planes when we open any one direction of $x/y$ directions. When the Chern numbers are even and any plane is not TSC, the Hamitonians with planes $k_{z}=0$ and $k_{z}=\pi$ are more like Hamitonians for topological insulators, and the two modes linked by the flat bands seem more like topological insulators' boundary states in the counter planes.

	\subsection{The Gapless systems}\label{Sec7}
\begin{table*}
\scriptsize
\caption{\label{tab1}The classification of topological gapless nodes in 3D systems. The first two rows denote the topological properties of the $k_{z}$=0 and $k_{z}=\pi$ plane. T, W and S mean trivial, WTSC and STSC, respectively. The third row denotes the values of $M_{3D}$, and the fourth row denotes the topological properties of the nodes in the 3D system. T, Dn and Wn mean trivial, Dirac and Weyl nodes, respectively. ``A/B" means that there is A or B, ``A (\&B)" means that there is A, or maybe there are both A and B. When $M_{3D}=-1$, there must be Weyl nodes in the 3D systems.}
\begin{tabular}{c|ccc|ccc|ccc}
\hline
$k_{z}=0$& &T & & &W & & & S& \\
\hline
$k_{z}=\pi$&T& W& S&T& W& S&T& W& S\\
\hline
$M_{3D}$&+1&+1&-1&+1&+1&-1&-1&-1&+1\\
\hline
The properties &\multirow{3}{*}{Dn/Wn}&  \multirow{3}{*}{Dn/Wn}&  \multirow{3}{*}{Wn(\& Dn)} &\multirow{3}{*}{Dn/Wn}&  \multirow{3}{*}{Dn/Wn}&  \multirow{3}{*}{Wn(\& Dn)} & \multirow{3}{*}{Wn(\& Dn)} & \multirow{3}{*}{Wn(\& Dn)} &  \multirow{3}{*}{Dn/Wn}\\
of the& &&&&&&&&\\
 Gapless nodes&&&&&&&&&\\
\hline
\end{tabular}
\end{table*}
The most interesting achievements we obtain in 3D systems in this work are that we can easily find whether a SC is gapless or not, and we can provide a simple method to search for Weyl superconductors by the $\mathbf{Z}_{2}$ invariant. We define the 3D $\mathbf{Z}_{2}$ invariant
\begin{eqnarray}\label{eq15}
M_{3D}=\prod_{k_{i}=0,\pi}M_{2D}(k_{i})=Sgn\{\prod_{i=1}^{8}Pf[B_{m}(K_{i})]\},
\end{eqnarray}	
where $M_{2D}(k_{i})$ refer to the $\mathbf{Z}_{2}$ invariant of counter plane $k_{i}=0$ and $k_{i}=\pi$, and $K_{i}$ are particle-hole symmetric invariant points.
In the gapful systems, $M_{3D}=+1$. However, when $M_{3D}=+1$, the systems do not must be gupful.
There are three cases with $M_{3D}=+1$. $Case$ 1: the two counter planes ($k_{i}=0,k_{i}=\pi$) are in different $\mathbf{Z}_{2}$ TSC phases; $Case$ 2: the two counter planes are in the same $\mathbf{Z}_{2}$ TSC phase with Chern numbers also being the same; $Case$ 3: the two counter planes are in the same $\mathbf{Z}_{2}$ TSC phase, but Chern numbers of the two planes differ by even numbers.
For case 1 and 3, the systems must be gapless, and while for case 2, the systems can be gapful or gapless because it is unclear whether the Chern number changes or not between the two planes. So, we can not determine whether the systems are gapful or gapless only by knowing $M_{3D}=+1$.

However, we do know that the systems must be gapless when $M_{3D}=-1$. In such systems, there must be gap closing between the two counter planes and the nodes are protected by the topological invariant $M_{3D}$. Assuming that the two counter planes are gapful, when $M_{3D}=-1$, the $M_{2D}$ of two counter planes must be different, and thus the two planes are in different topological phases. The phase changing between the two planes must accompany with gap closing.
We can also explain the nodes from the changing of Chern numbers. In the situation where the two counter planes are gapful, when $M_{3D}=-1$, the Chern numbers of the two counter planes must be different, and then the Chern number changes between the two planes, which is accompanied with gap closing. Note that the two counter planes may be not simultaneously gapful. If the plane is gapless, the definition of Chern number is invalid while the $\mathbf{Z}_{2}$ invariant is still valid. When one plane is trivial and gapful, its $M_{2D}$ equals +1. On the other hand, when the plane is topologically trivial and the invariant $M_{2D}$ is -1, the plane must be gapless. So the existence of $M_{3D}=-1$ also indicates that the system is gapless. We can conclude that, no matter whether the counter planes are gapful or gapless, $M_{3D}=-1$ can always be the proof of gap closing in the system.

We present the possible gapless situations in table \ref{tab1}.
The table shows that when one of the two planes is STSC, no matter whether the other one plane is trivial or WTSC, $M_{3D}=-1$ holds and Weyl nodes exist in the system. In such situations, the two counter planes are in different $\mathbf{Z}_{2}$ phases, and thus odd numbers of times of gap closing occur between the two planes. No matter whether the nodes are degenerate or not, there must be one non-degeneracy Weyl node. When one of the counter planes is WTSC and the other is trivial, we can not judge whether the nodes are Weyl or Dirac ones only by the Chern numbers or invariant $M_{2D}$ of the two planes, as we show in table \ref{tab1}.
When both planes are WTSC, the Chern numbers of them are even, and the nodes between the two planes are either even numbers of non-degenerate Weyl nodes or any numbers of Dirac nodes. However, we confirm that when $M_{3D}$=-1, the 3D system must have Weyl node and the system is Weyl superconductor.
Taking the model (\ref{eq13}) for example,
we write down the pfaffians of $k_{z}=\pi$ plane as
\begin{eqnarray}\label{eq16}
&&Pf[B_{m}(0,0)]=(\mu+2)^{2}-V^{2},\nonumber \\
&&Pf[B_{m}(0,\pi)]=Pf[B_{m}(\pi,0)]=\mu^{2}-V^{2},\nonumber\\
&&Pf[B_{m}(\pi,\pi)]=(\mu-2)^{2}-V^{2}.
\end{eqnarray}
Combining them with Eq. (\ref{eq14}), we get various gapless states by tuning $t_{z}$ and $\mu$, some of which are shown in Fig. (\ref{fig3})(b)-(f). We observe $M_{3D}=+1$ in Fig. (\ref{fig3})(b),(d)-(f) and $M_{3D}=-1$ in Fig. (\ref{fig3})(c).
	\section{Summary}\label{Sec8}
	In 1D TSCs, the Zak phase is equivalent to the Kitaev $\mathbf{Z_{2}}$ invariant. In 2D system without time reversal symmetry, the TSCs are suggested to characterized by $\mathbb{Z}$ invariant. We find the $\mathbb{Z}$ invariant can not tell us where the MZMs locate at and sometimes mismatch with the number of topological boundary states. We extend the Zak phase related $\mathbf{Z_{2}}$ invariant to 2D situation, and prove that it is identical to the parity of Chern number. The TSCs are distinguished as strong TSCs and weak TSCs with $\mathbf{Z_{2}}$ invariant $(M_{2D};M_{1},M_{2})$. When $M_{2D}=-1$, there are strong TSCs; when $M_{2D}=+1$, and either of $M_{i}$ is -1 or both of $M_{i}$ are -1, there are weak TSCs. We find that the $\mathbf{Z_{2}}$ invariants account for the mismatchness of the Chern numbers and Majorana zero modes and fix the location of the Majorana zero modes. We extend the $\mathbf{Z_{2}}$ invariant to 3D systems in which there have been no topological invariant by traditional topological classification. In gapful system, the 2D $\mathbf{Z_{2}}$ invariant of any one of the six boundary planes of the cubic BZ reflects the whole properties of the system. There are flat bands penetrating the 2D BZ to link the zero modes when the system is subject to the open boundary. The 3D $\mathbf{Z_{2}}$ invariant can be also used in gapless systems. When the 3D $\mathbf{Z_{2}}$ invariant is -1, the systems must be gapless and there must be Weyl nodes in the systems. One can take the 3D $\mathbf{Z_{2}}$ invariant as a simple criterion to identify whether a SC system is Weyl TSC or not.

    \section{Acknowledgments}\label{Sec9}
	This work was supported by NSFC under grant No.11947082, No.11975110, No.12065011, the Scientific and Technological Research Fund of Jiangxi Provincial Education Department under grant No. GJJ190577, the Zhejiang Provincial Natural Science Foundation of China (Grant No. LY21A050002), and the Research Start-up Fund from the Zhejiang Sci-Tech University(Grant No. 20062318-Y).


\begin{thebibliography}{0}%
\makeatletter
\providecommand \@ifxundefined [1]{%
 \@ifx{#1\undefined}
}%
\providecommand \@ifnum [1]{%
 \ifnum #1\expandafter \@firstoftwo
 \else \expandafter \@secondoftwo
 \fi
}%
\providecommand \@ifx [1]{%
 \ifx #1\expandafter \@firstoftwo
 \else \expandafter \@secondoftwo
 \fi
}%
\providecommand \natexlab [1]{#1}%
\providecommand \enquote  [1]{``#1''}%
\providecommand \bibnamefont  [1]{#1}%
\providecommand \bibfnamefont [1]{#1}%
\providecommand \citenamefont [1]{#1}%
\providecommand \href@noop [0]{\@secondoftwo}%
\providecommand \href [0]{\begingroup \@sanitize@url \@href}%
\providecommand \@href[1]{\@@startlink{#1}\@@href}%
\providecommand \@@href[1]{\endgroup#1\@@endlink}%
\providecommand \@sanitize@url [0]{\catcode `\\12\catcode `\$12\catcode
  `\&12\catcode `\#12\catcode `\^12\catcode `\_12\catcode `\%12\relax}%
\providecommand \@@startlink[1]{}%
\providecommand \@@endlink[0]{}%
\providecommand \url  [0]{\begingroup\@sanitize@url \@url }%
\providecommand \@url [1]{\endgroup\@href {#1}{\urlprefix }}%
\providecommand \urlprefix  [0]{URL }%
\providecommand \Eprint [0]{\href }%
\providecommand \doibase [0]{http://dx.doi.org/}%
\providecommand \selectlanguage [0]{\@gobble}%
\providecommand \bibinfo  [0]{\@secondoftwo}%
\providecommand \bibfield  [0]{\@secondoftwo}%
\providecommand \translation [1]{[#1]}%
\providecommand \BibitemOpen [0]{}%
\providecommand \bibitemStop [0]{}%
\providecommand \bibitemNoStop [0]{.\EOS\space}%
\providecommand \EOS [0]{\spacefactor3000\relax}%
\providecommand \BibitemShut  [1]{\csname bibitem#1\endcsname}%
\let\auto@bib@innerbib\@empty
\end{thebibliography}%


\begin{thebibliography}{99}
\bibitem{Kitaev2001}A. Y. Kitaev, Physics-Uspekhi 44, 131 (2001).
\bibitem{Read2000}N. Read and D. Green, Phys. Rev. B 61, 10267 (2000).
\bibitem{Ivanov2001}D. A. Ivanov, Phys. Rev. Lett. 86, 268 (2001).
\bibitem{Stone2006}M. Stone and S.-B. Chung, Phys. Rev. B 73, 014505 (2006).
\bibitem{Kitaev2003}A. Kitaev, Annals of Physics 303, 2 (2003).
\bibitem{Stern2004}A. Stern, F. von Oppen, and E. Mariani, Phys. Rev. B 70, 205338 (2004).
\bibitem{Das2006}S. Das Sarma, C. Nayak, and S. Tewari, Phys. Rev. B 73, 220502(R) (2006).
\bibitem{Nayak2008}C. Nayak, S. H. Simon, A. Stern, M. Freedman, and S. Das Sarma, Rev. Mod. Phys. 80, 1083 (2008).
\bibitem{Sau2012}J. D. Sau and S. D. Sarma, Nature Communications 3, 964 (2012).
\bibitem{Guang2016}G.-Y. Yi, X.-Q. Wang, Z. Gao, H.-N. Wu, and W.-J. Gong, Physica E: Low-dimensional Systems and Nanostructures 83, 481 (2016).
\bibitem{Zhang2016}P. Zhang and F. Nori, New Journal of Physics 18, 043033 (2016).
\bibitem{Kulesh2020}I. Kulesh, C. T. Ke, C. Thomas, S. Karwal, C. M. Moehle, S. Metti, R. Kallaher, G. C. Gardner, M. J. Manfra, and S. Goswami, Phys. Rev. Applied 13, 041003(R) (2020).
\bibitem{Estrada2020}J. C. Estrada Saldana, A. Vekris, R. Zitko, G. Steffensen, P. Krogstrup, J. Paaske, K. Grove-Rasmussen, and J. Nygard, Phys. Rev. B 102, 195143 (2020).
\bibitem{Scherubl2020}Z. Scherbl, G. Flp, C. P. Moca, J. Gramich, A. Baumgartner, P. Makk, T. Elalaily, C. Schnenberger, J. Nygrd, G. Zarnd, and S. Csonka, Nature Communications 11, 1834 (2020).
\bibitem{Hor2010}Y. S. Hor, A. J. Williams, J. G. Checkelsky, P. Roushan, J. Seo, Q. Xu, H. W. Zandbergen, A. Yazdani, N. P. Ong, and R. J. Cava, Phys. Rev. Lett. 104, 057001 (2010).
\bibitem{Wang2012}M.-X. Wang, C. Liu, J.-P. Xu, F. Yang, L. Miao, M.-Y. Yao, C. L. Gao, C. Shen, X. Ma, X. Chen, Z.-A. Xu, Y. Liu, S.-C. Zhang, D. Qian, J.-F. Jia, and Q.-K. Xue, Science 336, 52 (2012).
\bibitem{Matano2016}K. Matano, M. Kriener, K. Segawa, Y. Ando, and G.-q. Zheng, Nature Physics 12, 852 (2016).
\bibitem{Greco2018}A. Greco and A. P. Schnyder, Phys. Rev. Lett. 120, 177002 (2018).
\bibitem{Sun2019}Y. Sun, S. Kittaka, T. Sakakibara, K. Machida, J. Wang, J. Wen, X. Xing, Z. Shi, and T. Tamegai, Phys. Rev. Lett. 123, 027002 (2019).
\bibitem{Das2020}D. Das, K. Kobayashi, M. P. Smylie, C. Mielke, T. Takahashi, K. Willa, J.-X. Yin, U.Welp, M. Z. Hasan, A. Amato, H. Luetkens, and Z. Guguchia, Phys. Rev. B 102, 134514 (2020).
\bibitem{Adam2021}M. L. Adam, Z. Liu, O. A. Moses, X. Wu, and L. Song, Nano Research 14, 2613 (2021).
\bibitem{Fu2008}L. Fu and C. L. Kane, Phys. Rev. Lett. 100, 096407 (2008).
\bibitem{Williams2012}J. R. Williams, A. J. Bestwick, P. Gallagher, S. S. Hong, Y. Cui, A. S. Bleich, J. G. Analytis, I. R. Fisher, and D. Goldhaber-Gordon, Phys. Rev. Lett. 109, 056803 (2012).
\bibitem{Xu2014}J.-P. Xu, C. Liu, M.-X. Wang, J. Ge, Z.-L. Liu, X. Yang, Y. Chen, Y. Liu, Z.-A. Xu, C.-L. Gao, D. Qian, F.-C. Zhang, and J.-F. Jia, Phys. Rev. Lett. 112, 217001 (2014).
\bibitem{Qing2017}Q. L. He, L. Pan, A. L. Stern, E. C. Burks, X. Che, G. Yin, J. Wang, B. Lian, Q. Zhou, E. S. Choi, K. Murata, X. Kou, Z. Chen, T. Nie, Q. Shao, Y. Fan, S.-C. Zhang, K. Liu, J. Xia, and K. L. Wang, Science 357, 294 (2017).
\bibitem{Trang2020}N. K. S. S. S. K. W. I. Y. K. O. T. S. K. T. T. Y. A. Trang C. X., Shimamura N. and S. T., Nature Communications 11, 159 (2020).
\bibitem{Guo2021}L. Guo, Y. Yan, R. Xu, J. Li, and C. Zeng, Phys. Rev. Lett. 126, 057701 (2021).
\bibitem{Martin2012}I. Martin and A. F. Morpurgo, Phys. Rev. B 85, 144505 (2012).
\bibitem{Vazifeh2013}M. M. Vazifeh and M. Franz, Phys. Rev. Lett. 111, 206802 (2013).
\bibitem{Xiao2015}J. Xiao and J. An, New J. Phys. 17, 113034 (2015).
\bibitem{Pershoguba2016}S. S. Pershoguba, S. Nakosai, and A. V. Balatsky, Phys. Rev. B 94, 064513 (2016).
\bibitem{Li2016}J. Li, T. Neupert, B. A. Bernevig, and A. Yazdani, Nature Communications 7, 10395 (2016).
\bibitem{Howon2018}H. Kim, A. Palacio-Morales, T. Posske, L. Rzsa, K. Palots, L. Szunyogh, M. Thorwart, and R. Wiesendanger, Science Advances 4, eaar5251 (2018).
\bibitem{Rex2019}S. Rex, I. V. Gornyi, and A. D. Mirlin, Phys. Rev. B 100, 064504 (2019).
\bibitem{Choi2019}D.-J. Choi, N. Lorente, J. Wiebe, K. von Bergmann, A. F. Otte, and A. J. Heinrich, Rev. Mod. Phys. 91, 041001 (2019).
\bibitem{Kamlapure2021}A. Kamlapure, L. Cornils, R. itko, M. Valentyuk, R. Mozara, S. Pradhan, J. Fransson, A. I. Lichtenstein, J. Wiebe, and R. Wiesendanger, Nano Letters 21, 6748 (2021).
\bibitem{Stevan2014}S. Nadj-Perge, I. K. Drozdov, J. Li, H. Chen, S. Jeon, J. Seo, A. H. MacDonald, B. A. Bernevig, and A. Yazdani, Science 346, 602 (2014).
\bibitem{Rontynen2015}J. Rontynen and T. Ojanen, Phys. Rev. Lett. 114, 236803 (2015).
\bibitem{Heimes2015}A. Heimes, D. Mendler, and P. Kotetes, New Journal of Physics 17, 023051 (2015).
\bibitem{Pawlak2016}R. Pawlak, M. Kisiel, J. Klinovaja, T. Meier, S. Kawai, T. Glatzel, D. Loss, and E. Meyer, npj Quantum Information 2, 16035 (2016).
\bibitem{Ruby2017}M. Ruby, B. W. Heinrich, Y. Peng, F. von Oppen, and K. J. Franke, Nano Letters 17, 4473 (2017).
\bibitem{Zhang2020}G. Zhang, T. Samuely, N. Iwahara, J. Kamark, C. Wang, P. W. May, J. K. Jochum, O. Onufriienko, P. Szab, S. Zhou, P. Samuely, V. V. Moshchalkov, L. F. Chibotaru, and H.-G. Rubahn, Science Advances 6, eaaz2536 (2020).
\bibitem{Wada2021}K. Wada, T. Sugimoto, and T. Tohyama, Phys. Rev. B 104, 075119 (2021).
\bibitem{Wang2021}D. Wang, J. Wiebe, R. Zhong, G. Gu, and R. Wiesendanger, Phys. Rev. Lett. 126, 076802 (2021).
\bibitem{Schnyder2008}A. P. Schnyder, S. Ryu, A. Furusaki, and A. W. W. Ludwig, Phys. Rev. B 78, 195125 (2008).
\bibitem{Ryu2010}S. Ryu, A. P. Schnyder, A. Furusaki, and A. W. W. Ludwig, New Journal of Physics 12, 065010 (2010).
\bibitem{Qi2010}X. L. Qi, T. L. Hughes, and S.-C. Zhang, Phys. Rev. B 81, 134508 (2010).
\bibitem{Zhang2013}F. Zhang, C. L. Kane, and E. J. Mele, Phys. Rev. Lett. 111, 056403 (2013).
\bibitem{Ueno2013}Y. Ueno, A. Yamakage, Y. Tanaka, and M. Sato, Phys. Rev. Lett. 111, 087002 (2013).
\bibitem{Ando2015}Y. Ando and L. Fu, Annual Review of Condensed Matter Physics 6, 361 (2015).
\bibitem{Shiozaki2016}K. Shiozaki, M. Sato, and K. Gomi, Phys. Rev. B 93, 195413 (2016).
\bibitem{Wang2016Q}Q.-Z. Wang and C.-X. Liu, Phys. Rev. B 93, 020505(R) (2016).
\bibitem{Liu2017}X.-P. Liu, Y. Zhou, Y.-F. Wang, and C.-D. Gong, New Journal of Physics 19, 093018 (2017).
\bibitem{Zhang2018}R.-X. Zhang and C.-X. Liu, Phys. Rev. Lett. 120, 156802 (2018).
\bibitem{Zhang2019}R.-X. Zhang, W. S. Cole, and S. Das Sarma, Phys. Rev. Lett. 122, 187001 (2019).
\bibitem{Pan2019}X.-H. Pan, K.-J. Yang, L. Chen, G. Xu, C.-X. Liu, and X. Liu, Phys. Rev. Lett. 123, 156801 (2019).
\bibitem{Hu2020}L.-H. Hu, P. D. Johnson, and C. Wu, Phys. Rev. Research 2, 022021(R) (2020).
\bibitem{Ono2021}S. Ono, H. C. Po, and K. Shiozaki, Phys. Rev. Research 3, 023086 (2021).
\bibitem{Chiu2013}C.-K. Chiu, H. Yao, and S. Ryu, Phys. Rev. B 88, 075142 (2013).
\bibitem{Shiozaki2014}K. Shiozaki and M. Sato, Phys. Rev. B 90, 165114 (2014).
\bibitem{Geier2020}M. Geier, P. W. Brouwer, and L. Trifunovic, Phys. Rev. B 101, 245128 (2020).
\bibitem{Liu20172}F. Liu and K.Wakabayashi, Phys. Rev. Lett. 118, 076803 (2017).
\bibitem{Wu2020}H. C.Wu, L. Jin, and Z. Song, Phys. Rev. B 102, 035145 (2020).
\bibitem{Budich2013}J. C. Budich and E. Ardonne, Phys. Rev. B 88, 075419 (2013).
\bibitem{Fu2007}L. Fu, C. L. Kane, and E. J. Mele, Phys. Rev. Lett. 98, 106803 (2007).
\end{thebibliography}
\end{document}